\documentclass{article}
\usepackage{spconf}
\usepackage{amsmath}
\usepackage{graphicx, graphics, theorem, times, amsfonts, amsmath, amssymb, cite}
\usepackage{tikz}
\usetikzlibrary{shapes,arrows}
\usepackage{pgfplots}
\usepackage{color}
\usepackage{hyperref}
\usepackage{multirow}
\usepackage{rotating}
\usepackage{cleveref, subcaption}
\usepackage{bm}
\input{mysymbol.sty}
\usepackage[utf8]{inputenc}
\usepackage{enumitem}
\usepackage[linesnumbered]{algorithm2e}
\usepackage{array}
\newcolumntype{P}[1]{>{\centering\arraybackslash}p{#1}}

\newcommand{\QED}{\hfill\ensuremath{\blacksquare}}

\DeclareMathOperator{\vvec}{vec}

\ninept

\title{Graph Learning from Gaussian and Stationary Graph Signals}
\name{Andrei Buciulea and Antonio G. Marques
\thanks{Work supported by the Spanish NSF Grants SPGraph (PID2019-105032GB-I00/AEI/10.13039/501100011033) and DATRASCOOP@SESM (TED2021-130347B-I00), by the Young Researchers R\&D Project, ref. num. F861 (AUTO-BA- GRAPH) funded by the Community of Madrid (CAM) and Rey Juan Carlos University (URJC), and by the Grants F649-1209 and PREDOC20-003 funded by the CAM and URJC. All the authors are with the Dept. of Signal Theory and Comms., King Juan Carlos University, Madrid, Spain. Email contact author: antonio.garcia.marques@urjc.es.}}
\address{ Dept. of Signal Theory and Communications, King Juan Carlos University, Madrid, Spain}

\begin{document}
%\ninept
\maketitle
\begin{abstract}
Graphs have become pervasive tools to represent information and datasets with irregular support. However, in many cases, the underlying graph is either unavailable or naively obtained, calling for more advanced methods to its estimation. Indeed, graph topology inference methods that estimate the network structure from a set of signal observations have a long and well established history. By assuming that the observations are both Gaussian and stationary in the sought graph, this paper proposes a new scheme to learn the network from nodal observations. Consideration of graph stationarity overcomes some of the limitations of the classical Graphical Lasso algorithm, which is constrained to a more specific class of graphical models. On the other hand, Gaussianity allows us to regularize the estimation, requiring less samples than in existing graph stationarity-based approaches. While the resultant estimation (optimization) problem is more complex and non-convex, we design an alternating convex approach able to find a stationary solution. Numerical tests with synthetic and real data are presented, and the performance of our approach is compared with existing alternatives.

\end{abstract}
\section{Introduction}\label{sec:introudction}
Data defined over non-Euclidean supports is nowadays ubiquitous, with graphs, due to their versatility and tractability, having emerged as a suitable alternative to represent those irregular domains. This has prompted the emergence of works investigating how to generalize classical information processing architectures to graph-based signals. The effort has permeated also to applications, with a number of graph-based data-processing schemes being successfully applied to electrical, communication, social, geographic, financial, genetic, and brain networks ~\cite{kolaczyk2009book,sporns2012book,EmergingFieldGSP,ortega_2018_graph,alglearninggraphs}, to name a few.

When dealing with network data, the workhorse assumption is that the underlying graph is known. Unfortunately, this is not always the case. 
In many scenarios, the graph structure is not well defined, either because there is no underlying physical network (e.g., when the graph models logical relations between variables) or because there is no unique metric to measure the relationship between the different nodes of the network.  Since in most cases such relationships cannot be known, the graph topology is usually inferred from the set of available observations.
To estimate the graph topology, a first step is to formalize the relation between the network edges and the properties of the signals that live in the network.
The early graph topology inference methods~\cite{mateos2019connecting,sardellitti2019graph} adopted a statistical approach, such as the correlation network~\cite[Ch. 7.3.1]{kolaczyk2009book}, partial correlations or Gaussian Markov Random Fields (MRFs)~\cite{meinshausen06,kolaczyk2009book}, with graphical Lasso (GL)~\cite{GLasso2008} being the workhorse scheme for their estimation. In recent years, graph signal processing (GSP) based models exploiting properties such as smoothness or graph stationarity~\cite{MeiGraphStructure,egilmez2017graph,segarra2017network,buciulea2021learning}  have become increasingly important, along with their nonlinear generalizations\cite{Karanikolas_icassp16}.
The main advantage of heuristic pairwise distance-based and correlation-based approaches, which infer each link separately, is that the graph can be learned even in setups where only a few observations are available. In contrast, theoretically sounded methods that postulate a more advanced relation (mapping) between the observations and the totality of the graph require a significantly larger number of observations. Remarkably, GL achieves a sweet spot in this regard, postulating a relatively sophisticated model (a Gaussian MRF) and being able to learn the graph using only a moderate number of observations. 

In this context, this paper proposes a new graph-learning method to estimate a network from a set of signal observations. Our main contributions are: i) building a new graph-learning approach around the assumption that the observations at hand are both Gaussian and stationary in the graph; ii) formulating a joint optimization problem that yields as a result the sough graph along with an enhanced estimate of the precision matrix of the Gaussian process; and iii) designing an efficient provably convergent algorithm to address the (non-convex) optimization in ii). Relative to GL, which assumes Gaussianity but not stationarity, we end up with a more general graph learning algorithm able to recover the graph in a wider range of scenarios and that includes GL as a particular case.
For our problem formulation we made assumptions related  to 1) the graph (to be sparse), 2) the signals (to be Gaussian) and 3) the relationship between the graph and the signals (to be graph stationary). 
Relative to graph-stationary-based approaches, the scheme in this paper is able to work with less samples, is more robust to noise and perturbations, and can be used to draw links with classical statistical approaches. The price to pay is that the resultant algorithm incurs higher computational complexity and that, due to non-convexity, only convergence to a stationary point is guaranteed. Interestingly, the numerical results presented, while limited, show that our algorithm either outperforms or matches the results achieved by its convex counterparts.

\noindent\textbf{{Outline.}} The remaining of the paper is organized as follows. Sec.~\ref{S:fundamental_GSP} surveys the required graph background, making special emphasis on the concepts of Gaussian and graph stationary signals.
In Sec.~\ref{S:netw_reconstruction} we formalize the problem of network topology inference from signals that are Gaussian and graph stationary. Sec.~\ref{S:proposed_form} presents a bi-convex optimization formulation of the problem in Sec. \ref{S:netw_reconstruction} and develops a provably convergent alternating iterative algorithm to solve it. Sec.~\ref{S:numerical_experiments} illustrates the recovery performance of the proposed algorithm using both synthetic and real data. %Concluding remarks in Sec.~\ref{S:conclusions} close the paper.

%%%%%%%%%%%%%%%%%%%%%%%%%%%%%%%%%%%%%%%%%%%%%%%%%%%%%%%%%%%%%%%%%%%%%%%%%

\section{Fundamentals of graph signal processing} \label{S:fundamental_GSP}
This section introduces GSP concepts that help to formalize the learning problem and explain the relation between the available signals and the topology of the underlying graph. 
    
\vspace{0.1cm}
\noindent\textbf{GSO and graph signals.}
Let $\ccalG = \{\ccalV, \ccalE\}$ denote an \emph{undirected} and weighted graph with $N$ nodes where $\ccalV$ and $\ccalE$ represent the vertex and edge set, respectively.
For any $\ccalG$, its weighted adjacency matrix $\bbA\in\reals^{N \times N}$ is a sparse matrix with $A_{ij}$ capturing the weight of the edge between the nodes $i$ and $j$.
A general representation of the graph is the graph-shift operator GSO $\bbS \in \mathbb{R}^{N \times N}$, where $S_{ij} \neq 0$ if and only if $i=j$ or $(i,j) \in \ccalE$ \cite{ortega_2018_graph}.
Typical choices for the GSO are the adjacency matrix $\bbA$, the combinatorial graph Laplacian  $\bbL := \diag(\bbA \textbf{1}) - \bbA$, and their respective generalizations.
The signals defined on the nodes of the graph $\ccalG$ are called graph signals. A graph signal can be denoted as a vector $\bbx\in \mathbb{R}^{N}$ with $x_{i}$ being the signal value observed at node $i$.

\vspace{0.1cm}
\noindent\textbf{Gaussian graph signals.}
Since graph signals can be represented by $N$-dimensional vectors, the random (zero-mean) graph signal $\bbx$ 
%\red{where $f_{\bbTheta}(\bbx)$ is the Gaussian joint distribution cite},
is said to be Gaussian, if its joint distribution has the form 
%\red{Para ahorrar espacio lo que haría sería eliminar una de estas ecuaciones (la eq. 1, poniendo una referencia por ejemplo) o tratar de juntarlas para que ocupen menos}
\vspace{-0.3cm}
\begin{equation}\label{E:G_sig}
    f_{\bbTheta}(\bbx) = {(2\pi)^{-N/2}} \cdot \det(\bbTheta)^{\frac{1}{2}} \cdot \exp\big({-\frac{1}{2}\bbx^{T}\bbTheta\bbx}\big).
    \vspace{-0.2cm}
\end{equation}
with $\bbTheta$ being the precision matrix  (the inverse of the covariance matrix $\bbSigma$). Suppose now that we have a collection of $R$ Gaussian signals $\bbX = [\bbx_1,...,\bbx_R] \in \mathbb{R}^{N \times R}$ each of them independently drawn from the distribution in \eqref{E:G_sig}. The log-likelihood associated with $\bbX$ is
\vspace{-0.2cm}
\begin{equation}\label{E:Like}
    L(\bbX|\bbTheta) = \prod_{r=1}^{R} f_{\bbTheta}(\bbx_r),  \ \ \ \ccalL(\bbX|\bbTheta) = \sum_{r=1}^{R} \log(f_{\bbTheta}(\bbx_r)).
\vspace{-0.1cm}
\end{equation}
This expression will be exploited when formulating our proposed inference approach and establishing links with classical methods.

\vspace{0.1cm}
\noindent\textbf{Graph filters.}
A versatile tool to model the relationship between the signal $\bbx$ and its underlying graph are graph filters. 
%Since the graph is undirected, the GSO is symmetric and can be diagonalized as $\bbS=\bbV\bbLambda \bbV^\top$, where the diagonal matrix $\bbLambda$ collects the eigenvalues of the GSO and the orthogonal matrix $\bbV\in\reals^{N \times N}$ its eigenvectors.
A graph filter $\bbH\in\reals^{N\times N}$ is a linear operator that accounts for the topology of the graph and is defined as a polynomial of the GSO $\bbS$ of the form
\vspace{-0.2cm}
\begin{equation}\label{E:graph_filter}
    \bbH = \sum_{l=0}^{L-1} h_{l} \bbS^{l}= \sum_{l=0}^{L-1} h_{l}\bbV\bbLambda^l\bbV^\top=\bbV\Big(\sum_{l=0}^{L-1} h_{l}\bbLambda^l\Big)\bbV^\top,
    \vspace{-0.1cm}
\end{equation}

where $L-1$ is the filter degree, $\{h_l\}_{l=0}^{L-1}$ represent the filter coefficients, and $\bbV$ and $\bbLambda$ are the eigenvectors and eigenvalues of the GSO respectively.
Since $\bbH$ is a polynomial of $\bbS$, it readily follows that both matrices have the same eigenvectors $\bbV$.
\vspace{-0.1cm}

\vspace{1mm}
%Stationary signals and graph filters
\noindent\textbf{Stationary graph signals.} A random graph signal $\bbx$ is stationary in the GSO $\bbS$ if it can be represented as the output generated by a graph filter $\bbH$ [cf.\eqref{E:graph_filter}] whose input is a zero mean white signal $\bbw \in \mathbb{R}^{N}$ with $\mathbb{E}[\bbw\bbw^\top]=\bbI$. Under this definition, it follows that if $\bbx$ is stationary and $\ccalG$ is undirected, the  covariance of $\bbx$ is given by
\vspace{-0.1cm}
\begin{equation}\label{E:cov_x}
    \bbSigma\!=\! \mathbb{E}[\bbx\bbx^{\!\top}\!]\!=\!\mathbb{E}[\bbH\bbw\bbw^{\!\top}\bbH^{\!\top}]\!=\!\bbH\mathbb{E}[\bbw\bbw^{\!\top}\!]\bbH^{\!\top}\!=\!\bbH\bbH^{\!\top}\!=\!\bbH^2,
    \vspace{-0.1cm}
\end{equation}
where the last equality follows from the symmetry of the GSO. Since $\bbH$ is a polynomial in $\bbS$, so is $\bbSigma$. This implies that: i) $\bbTheta$ is also a polynomial of $\bbS$, and ii) $\bbS$, $\bbSigma$ and $\bbTheta$ share the same eigenvectors $\bbV$~\cite{marques2017stationary,perraudin2017stationary,Girault15}. As a byproduct, we also have that matrices $\bbS$ and $\bbSigma$ commute,
%i.e., $\bbSigma\bbS = \bbS\bbSigma$,
which are properties to be exploited later on. 
%\red{Andrei, saltamos de $\bbSigma$ a $\bbC$ esto no puede ser. Elige la notacion que quieras, pero luego sé consistente.}
    
\vspace{-0.3cm}
\begin{table*}[]
%\caption{A}
\renewcommand{\arraystretch}{1.5}
\begin{align}
\vspace{-0.1cm}\bbTheta_1^{(t+1)} &=  \argmin_{\bbTheta_1\succeq 0}  \tr(\hbSigma\bbTheta_1) + \frac{\lambda_1}{2}     \|\bbTheta_1\bbS^{(t)}\!-\!\bbS^{(t)}\bbTheta_1\|_F^2 + \frac{\lambda_2}{2} \|\bbTheta_1\!-\!\bbTheta_2^{(t)}\|_F^2   \label{eq:StepI} \tag{10}\\
\vspace{-0.1cm} \bbTheta_2^{(t+1)} &= \argmin_{\bbTheta_2\succeq 0} \            -\log\det(\bbTheta_2)
              + \frac{\lambda_2}{2} \|\bbTheta_1^{(t+1)}-\bbTheta_2\|_F^2 \label{eq:StepII}\tag{11}\\ 
\vspace{-0.1cm} \bbS^{(t+1)} &= \argmin_{\bbS \in \ccalS} \ \rho\|\bbS\|_1  +                       \frac{\lambda_1}{2} \|\bbTheta_1^{(t+1)} \bbS-\bbS\bbTheta_1^{(t+1)}\|_F^2 \label{eq:StepIII}\tag{12}  
\vspace{-0.4cm}
\\
\hline \nonumber
\end{align}
\vspace{-.3cm}

%\begin{tabular}{|P{1.2cm}|P{15cm}|}
%\hline 
%\multicolumn{2}{|c|}{\textbf{Table 1:} BSUM graph-learning method for Gaussian
%and stationary signals (BSUM-GGSR)} \\
%\hline
%{\small \textbf{Step I}} & $\bbTheta_1^{(t+1)}=  \argmin_{\bbTheta_1\succeq 0} 
%             \tr(\hbSigma\bbTheta_1) + \frac{\lambda_1}{2}     \|\bbTheta_1\bbS^{(t)}\!-\!\bbS^{(t)}\bbTheta_1\|_F^2 + \frac{\lambda_2}{2} \|\bbTheta_1\!-\!\bbTheta_2^{(t)}\|_F^2 $ \\ \hline
%{\small\textbf{Step II}} & $\bbTheta_2^{(t+1)}= \argmin_{\bbTheta_2\succeq 0} \
%            -\log\det(\bbTheta_2)
%              + \frac{\lambda_2}{2} \|\bbTheta_1^{(t+1)}-\bbTheta_2\|_F^2$ \\ \hline
%{\small\textbf{Step III}} & $\bbS^{(t+1)} = \argmin_{\bbS \in \ccalS} \ \rho\|\bbS\|_1  +                       \frac{\lambda}{1} \|\bbTheta_2^{(t+1)} \bbS-\bbS\bbTheta_2^{(t+1)}\|_F^2$ \\ \hline
%{\small\textbf{Dual updates}} & $\bbY^{(t+1)} = \bbY^{(t)} + \lambda \left( \bbTheta_2^{(t+1)}\bbS^{(t+1)}-\bbS^{(t+1)}\bbTheta_2^{(t+1)} \right), \ \ 
%\bbZ^{(t+1)} = \bbZ^{(t)} + \lambda \left( \bbTheta_1^{(t+1)}-\bbTheta_2^{(t+1)} \right)$ \\ \hline
%\end{tabular}
%\label{table:T1}
\vspace{-.6cm}
\end{table*}

%%%%%%%%%%%%%%%%%%%%%%%%%%%%%%%%%%%%%%%%%%%%%%%%%%%%%%%%%%%%%%%%%%%%%%%%%%
\section{Graph learning problem formulation}\label{S:netw_reconstruction}
%Topology inference problem definition
To formally state the graph learning problem, let $\bbX :=[\bbx_1,...,\bbx_R] \in \reals^{N \times R}$ denote a collection of $R$ independent realizations of a random process defined on the $N$ nodes of the unknown graph $\ccalG$. 
The goal of graph learning problems is to find the optimal graph descriptor (i.e., the GSO $\bbS$ associated with $\ccalG$) given the set of signals $\bbX$.
There exist a range of approaches in the literature to address this problem~\cite{GLasso2008, dong_2019_learning, marques2017stationary}. In this paper, we approach it by assuming that the signals in $\bbX$ are \emph{simultaneously} Gaussian and stationary in the sought graph. Based on these two assumptions (along with the standard consideration of the graph being sparse), we propose a new approach more robust and general than those existing in the literature, able to recover the graph in a wider range of scenarios. This section starts by formalizing the problem we want to address and then, formulates it as an optimization problem. After that, we discuss the main features of our approach and elaborate in detail on the differences between GL and graph-stationary approaches. 
%Our proposal statement

\vspace{0.1cm}
\noindent\textbf{Problem 1.} \textit{Given the set of signals $\bbX \in \reals^{N \times R}$, find the underlying sparse graph structure encoded in $\bbS$ under the assumptions:} 

\noindent \textit{(AS1): The graph signals in $\bbX$ are i.i.d. realizations of $\ccalN(\bb0, \bbSigma)$}

\noindent\textit{(AS2): The graph signals in $\bbX$ are stationary in $\bbS$.}
\vspace{0.1cm}

\noindent Our approach is to recast Problem 1 as the following optimization 
\vspace{-0.2cm}
\begin{alignat}{2}\label{E:alg_GGSR}
    \!\!&\!\hbTheta, \hbS = \argmin_{\bbTheta\succeq 0,\bbS \in \ccalS} \ \nonumber
    && -\log(\det(\bbTheta)) + \tr(\hbSigma\bbTheta) + \rho \|\bbS\|_0.\\
        \!\!&\!\mathrm{\;\;s. \;to } &&   \bbTheta\bbS=\bbS\bbTheta, \tag{5}
        \vspace{-0.6cm}
\end{alignat}
where $\ccalS$ is a generic set representing additional constraints that $\bbS$ is known to satisfy (e.g., the GSO being symmetric on its entries being between zero and one). The minimization takes as input the sample covariance matrix $\hbSigma=\frac{1}{R}\bbX\bbX^\mathsf{T}$ and generates as output the estimate for $\bbS$ and, as a byproduct, the estimate for $\bbTheta$. Next, we explain the motivation for each of the terms in \eqref{E:alg_GGSR} and, especially, for the constraint $\bbTheta\bbS=\bbS\bbTheta$, which is critical in our approach. The first two terms in the objective function come from leveraging (AS1) and are the result of minimizing the negative log-likelihood expression in \eqref{E:Like}. In contrast, the term $\rho \|\bbS\|_0$ accounts for the fact of $\bbS$ being a GSO (hence, sparse) with $\rho$ being a regularization constant / multiplier/ prior related to the sparsity level of the graph.
Finally, the equality constraint codifies (AS2). Note that the polynomial relation between $\bbSigma$ and $\bbS$ implied by (AS2) is typically codified in estimation/optimization problems by either: i) extracting the eigenvectors of $\bbSigma$ and imposing than those should be the eigenvectors of $\bbS$ \cite{segarra2017network} or ii) imposing the constraint $\bbSigma\bbS=\bbS\bbSigma$ \cite{buciulea2021learning}. Differently, here we codify the polynomial relation implied by (AS2) by imposing commutativity between $\bbTheta$ and $\bbS$. This is not only equivalent (if the matrices are full rank and $\bbSigma$ and $\bbS$ commute, the same is true for $\bbTheta$ and $\bbS$ --for more details check the Cayley-Hamilton theorem), but also more convenient when dealing with Gaussian signals, since their likelihood is convex with respect to $\bbTheta$.
It is important to note that while the assumption of stationarity may appear strong, it is in fact more lenient (more general) than an MRF, as it allows the precision matrix to be a polynomial function of $\bbS$. This stands in contrast to the GL framework, which limits this mapping to the equality constraint, $\bbS=\bbTheta$.
%For the last constraint, since we assume that $\hbSigma$ could be any polynomial in $\bbS$, $\bbTheta$ may not always be sparse. %For this reason, to promote sparsity in the graph we add a $\ell_0$ norm to $\bbS$ and use $\kappa$ as a regularizer to limit the maximum number of edges in the graph.

From a conceptual point of view, our formulation reaches a sweet spot between GL and graph-stationarity approaches. The price to pay is that the resultant algorithm (even if the $\ell_0$ norm is relaxed) is not convex due to the bilinear constraint coupling the \emph{optimization} variables $\bbTheta$ and $\bbS$. The remaining of this section provides additional details about the relation between \eqref{E:alg_GGSR}, GL and graph-stationary approaches, while Sec.~\ref{S:proposed_form} designs an efficient algorithm to deal with the non-convexities. 
%Relative to GL, our approach is more general than other alternatives because we assume that $\bbC$ can be any polynomial of $\bbS$. In contrast, GL assumes that $\bbS$ is given by $\bbC^{-1}$. Relatively to graph-stationary approches  graph can be recovered from a relatively small number of signals because of considering that the signals are Gaussian (AS1). To better understand the advantages of our algorithm, next, we will discuss briefly the implications of each of the assumptions separately.

Suppose first that we simplify (AS2) imposing that $\sigma\bbI+\delta\bbS=\bbSigma^{-1}$. 
Then, up to the diagonal values and scaling issues, the sparse matrix $\bbS$ to be estimated and $\bbTheta=\bbSigma^{-1}$ are the same and, as a result, it suffices to optimize over one of them. Upon defining $\ccalS_{\bbTheta} := \{ \bbTheta = \bbTheta^{T}; \ \bbTheta \succeq 0\}$, this leads to
\vspace{-0.3cm}
\begin{alignat}{2}\label{E:glasso}\tag{6}
      \!\!&\!\hbTheta= \argmin_{\bbTheta\succeq 0, \bbTheta \in \ccalS_{\bbTheta}} \
      && -\log(\det(\bbTheta)) + \tr(\hbSigma\bbTheta) + \rho \|\bbTheta\|_0,
      \vspace{-0.3cm}
\end{alignat}
which coincides exactly with the GL formulation \cite{GLasso2008}.
The main advantages of \eqref{E:glasso} relative to \eqref{E:alg_GGSR} are that the number of variables is smaller and the resultant problem (after relaxing the $\ell_0$ norm) is convex. The main drawback is that by forcing the support of $\bbS$ and $\bbTheta$ to be the same, the set of feasible graphs (and their topological properties) is more limited. Remarkably, when the model assumed in  \eqref{E:glasso} holds true (i.e., data is Gaussian and $\bbTheta$ is sparse), GL is able to find reliable estimates of $\bbS$ even when the number of samples $R$ is fairly low. On the other hand, simulations will show that GL does a poor job estimating $\bbS$ when the relation between the precision matrix and $\ccalG$ is more involved. 

Suppose now that we drop (AS1), so that we approach the graph learning problem assuming only that signals are stationary in $\ccalG$. 
Since $\bbSigma$ shares the same eigenvectors, if the covariance if full rank the expression $\bbTheta\bbS = \bbS \bbTheta$ is equivalent to requiring $\bbSigma\bbS = \bbS \bbSigma$. This allows us to rewrite \eqref{E:alg_GGSR} as 
\vspace{-0.1cm}
\begin{alignat}{1}\label{E:alg_GSR}\tag{7}
    \vspace{-0.4cm}
    \!\!&\!\hbS = \argmin_{\bbS \in \ccalS} \
     \rho \|\bbS\|_0  \; \; \; \mathrm{\;\;s. \;to }  \; \; \; \hbSigma\bbS=\bbS\hbSigma, 
     \vspace{-0.2cm}
\end{alignat}
where the constraint is typically relaxed as $\|\hbSigma\bbS-\bbS\hbSigma\|_F^2\leq \epsilon$ to account for the fact that we have $\hbSigma\approx \bbSigma$.
While more general than either \eqref{E:alg_GGSR} or \eqref{E:glasso}, the absence of Gaussianity implies that \eqref{E:alg_GSR} being able to identify the ground truth $\bbS$ requires very reliable estimates of $\hbSigma$, so that $\epsilon$ can be set close to zero. This is indeed a challenge, especially in setups when the number of nodes is large. 

%It can be seen that the proposed method is more complex and it is also non-convex due to the $\ell_0$ norm and the bi-linear term of the commutativity constraint. 
%In the next section we will deal with this issue and provide an iterative solution to problem \eqref{E:alg_GGSR}.
%%%%%%%%%%%%%%%%%%%%%%%%%%%%%%%%%%%%%%%%%%%%%%%%%%%%%%%%%%%%%%%%%%%%%%%%%%%%%%%%%%%%%%%%%
\vspace{-0.3cm}
\section{Graph learning algorithm}\label{S:proposed_form}
\vspace{-0.2cm}
The formulation in \eqref{E:alg_GGSR}, which optimizes jointly over $\bbTheta$ and $\bbS$, is non-convex due to the presence of the $\ell_0$ norm and the bilinear constraint. 
In this section, we design an algorithm to solve a relaxed version of \eqref{E:alg_GGSR} in an efficient way. The most important design steps include: i) relaxing the $\ell_0$ norm in the objective with the $\ell_1$ norm; ii) renaming variable $\bbTheta$ as $\bbTheta_1$, creating an auxiliary variable $\bbTheta_2$, and adding a new constraint that enforces $\bbTheta_1=\bbTheta_2$; iii) augmenting the cost with quadratic terms that account for the constraints; and iv) proposing a block coordinate descent framework that optimizes each of the blocks of variables ($\bbTheta_1$, $\bbTheta_2$, and $\bbS$) in an iterative manner.
%and iv) use a dual proximal gradient approach to deal with the linear constraints. 
%The proposed algorithm involves two main modifications: (i) the $\ell_0$ norm is replaced by the $\ell_1$ norm, which is its convex relaxation, and (ii) a new optimization variable $\bbTheta_2$ is added, to decouple the original problem which will allow us later to solve \eqref{E:alg_GGSR} in three simple steps. The resulting formulation after applying these modifications is as follows
Specifically, after incorporating steps i) and ii),  \eqref{E:alg_GGSR} is re-written as 
\vspace{-0.3cm}
\begin{alignat}{2}\label{E:alg_GGSR_convx} 
    \!\!&\!\hbTheta_1,\hbTheta_2, \hbS = && \!\!\!\!\! \argmin_{\bbTheta_1,\bbTheta_2\succeq 0;\bbS \in \ccalS} \
     -\log(\det(\bbTheta_1)) + \tr(\hbSigma\bbTheta_2) + \rho\|\bbS\|_1, \nonumber \\ 
    \!\!&\!\mathrm{\;\;s. \;to }  &&  \bbTheta_2\bbS=\bbS\bbTheta_2  \;\;\text{and}\;\; \bbTheta_1=\bbTheta_2. \tag{8}
    \vspace{-0.4cm}
\end{alignat}  

%%%%%%%%%%%%%%%%%%%
\RestyleAlgo{ruled}
\begin{algorithm}[tb]
    %\SetKwInOut{Input}{Input}
    \SetKwInOut{Output}{Outputs}
    \SetKwInput{KwData}{Input}
    \KwData{$\hbSigma$}
    \vspace{0.1cm}
    \Output{$\hbTheta_1$, $\hbTheta_2$, and $\hbS$}
    \SetAlgoLined
    \vspace{0.1cm}

    Initialize  $\hbTheta_2^{(0)}$, $\hbS^{(0)}$, $=0$.
    \\ \vspace{0.1cm}
    
    \While{not converged}{
    
    Update  $\hbTheta_1^{(t+1)}$ by solving \eqref{eq:StepI} (Step I). \\ \vspace{0.05cm}
    
    Update  $\hbTheta_2^{(t+1)}$ by solving \eqref{eq:StepII} (Step II). \\ \vspace{0.05cm}
    
    Update  $\hbS^{(t+1)}$ by solving \eqref{eq:StepIII} (Step III). \\ \vspace{0.05cm}
    
    %Update $\bbY^{(t+1)}$ and $\bbZ^{(t+1)}$ via \blue{Dual updates}.
    
    $t = t + 1$ \\ \vspace{0.05cm}
    
    }
    $\hbTheta_1=\hbTheta_1^{(t)}$,$\hbTheta_2=\hbTheta_2^{(t)}$,$\hbS=\hbS^{(t)}$
    \caption{BSUM graph-learning method for Gaussian and stationary signals (BSUM-GGSR)}
    \label{A:BSUM}
\end{algorithm}
\noindent Next, to deal with equality constraints $\bbTheta_2\bbS=\bbS\bbTheta_2$ and $\bbTheta_1=\bbTheta_2$, we augment the objective in \eqref{E:alg_GGSR_convx} with quadratic terms whose weights are controlled by regularization parameters  $\lambda_1$ and $\lambda_2$, yielding 
\vspace{-0.3cm}
\begin{alignat}{2}\label{E:GGSR_aug_lag}
     \!\!&\!\hbTheta_1,\hbTheta_2, \hbS = && \argmin_{\bbTheta_1,\bbTheta_2\succeq 0;\bbS \in \ccalS}  \tr(\hbSigma\bbTheta_1) - \log(\det(\bbTheta_2))  + \rho\|\bbS\|_1 \nonumber  \\  
    \!\!&\! && + \frac{\lambda_1}{2}\|\bbTheta_1\bbS-\bbS\bbTheta_1\|_F^2 
     + \frac{\lambda_2}{2}\|\bbTheta_1-\bbTheta_2\|_F^2. \tag{9} 
\end{alignat} 

\begin{figure*}
	\centering
    \includegraphics[height=4.3cm, width=18.1cm]{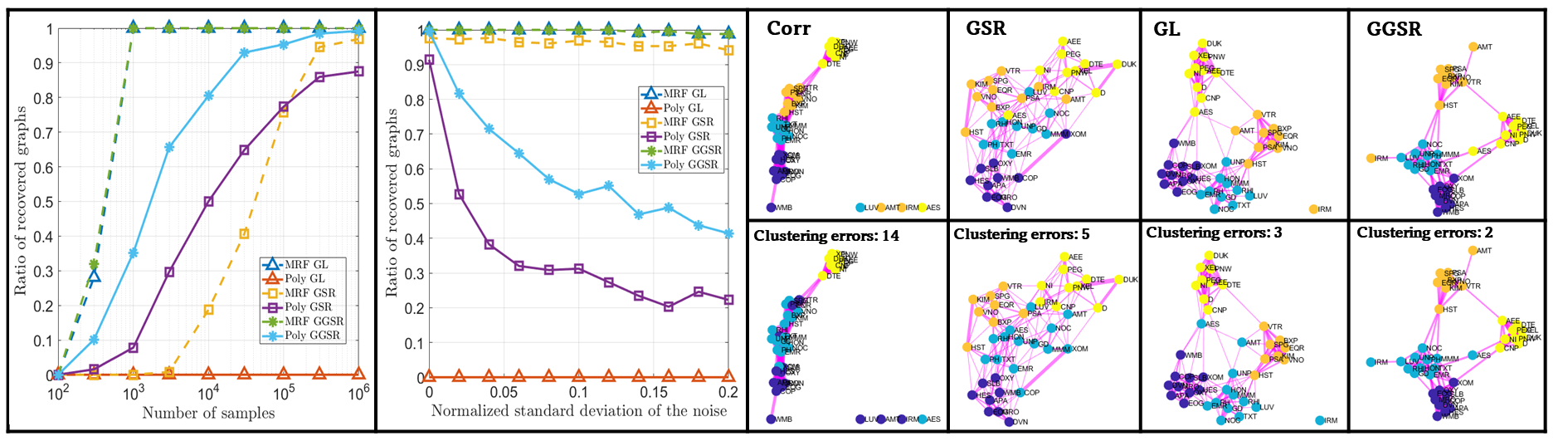}
	\caption{(Left) Ratio of recovered graphs 
	%($1$ if all the links are successfully recovered and $0$ otherwise) 
	while increasing the number of signals $R$ used to estimate $\hbSigma$. The values in the y-axis correspond to averages across $100$ realizations of Erdös-Rényi graphs with $N=20$ and link probability $p=0.1$. (Center) Setup similar to the previous one, but setting $R=10^6$ and with the x-axis representing the normalized power noise present in the observations. (Right) Visual representations of recovered graphs using as input financial data signals of 40 companies from 4 different sectors.}%\red{Faltaria mejorar la calidad de la figura o hacerla más grande}} 

	\label{F:num_exp}
\end{figure*}

\vspace{-0.3cm}

The pseudocode describing how to handle the optimization of (9) is detailed in Alg.~1. %, with the closed-form expression for each of the steps being provided in Table 1.
The overall idea is to solve the problem via a Block Successive Upper Bound Minimization (BSUM) alternating minimization approach for the three blocks of variables.
The iterative algorithm, referred to as graph Gaussian stationary recovery (GGSR), consists in successively optimizing the three blocks of variables with the other two blocks fixed.  
Note that, thanks to the auxiliary variables $\bbTheta_1$ and $\bbTheta_2$, the optimization for each of the three blocks is relatively simple and easy to solve (details are omitted due to the page limit constraints, but will be provided in the journal version of this work). 

Convergence guarantees of the iterative algorithm to a stationary point of function in (9) are formally stated next.

\vspace{0.1cm}
\noindent\textbf{Proposition 1.} \textit{Denote with $f$ the objective function in \eqref{E:GGSR_aug_lag}. 
Let $\ccalY^*$ be the set of stationary points of \eqref{E:GGSR_aug_lag}, and let $\bby^{(t)}=[\vvec(\bbTheta_1^{(t)})^\top\!,$ $\vvec(\bbTheta_2^{(t)})^\top\!,\vvec(\bbS^{(t)})^\top]^\top$ be the solution generated after running the three steps in Table 1 $t$ times.
Then, the solution generated by the iterative algorithm presented in Table 1 converges to a stationary point of $f$ as $t$ goes to infinity, i.e., 
}
\vspace{-0.3cm}
\[\lim_{t\to\infty} d(\bby^{(t)},\ccalY^*) = 0,\]
\textit{with $d(\bby,\ccalY^*) := \min_{\bby^* \in \ccalY^*} \|\bby-\bby^*\|_2$.}

The proof relies on the convergence results shown in ~\cite[Th. 1b]{hong2015unified}. 
It can be verified that our iterative algorithm meets the required conditions for the convergence to a stationary point, which are the following: 1) $f(\bby)$ is regular for all feasible points; 2) the first-order behavior of each step (when the remaining steps are fixed) is the same as $f(\bby)$; 3) the set $\ccalY^{(0)} = \{\bby | f(\bby) \leq f(\bby)^{(0)}\}$ is compact; and 4) Steps I and II have a unique solution.
Note that the convergence of the BSUM-GGSR algorithm was not obvious since the original function in \eqref{E:GGSR_aug_lag} is non-convex and Step III has no unique solution, we refer readers to \cite{shafipour2020online} for more details. 

The computational complexity required by Alg.~1 is polynomial but, in the worst case, it can scale as $O(N^7)$, which may hinder its application to large graphs. 
Although the implementation of more efficient and scalable algorithms is beyond the scope of this conference paper, we find it relevant and it is subject of our current work.
\vspace{-0.5cm}
%%%%%%%%%%%%%%%%%%%%%%%%%%%%%%%%%%%%%%%%%%%%%%%%%%%%%%%%%%%%%%%%%%%%%%%%%%
\vspace{-0.5cm}
\section{Numerical experiments}\label{S:numerical_experiments}
\vspace{-0.3cm}
This section numerically assesses the performance of our proposed algorithm, GGSR, comparing it with the two most related alternatives in literature: GL \cite{GLasso2008} and GSR \cite{segarra2017network}.
% General settings
Fig. \ref{F:num_exp} summarizes the results for synthetic data (left and center panels) and real data (right panel). 
For the synthetic experiments, the network topology is considered to be successfully recovered (score 1) if \textit{each and every} link is successfully estimated and zero otherwise. 
This score is averaged over $100$ realizations of random graphs with $N=20$ nodes following an Erdös-Rényi model with link probability $p=0.1$.
Regarding the generation of the graph signals, two different models for the covariance matrices have been studied: 1) $\bbC_{poly}=\bbH^{2}$ with $\bbH =\sum_{l=0}^L h_l \bbS^l$ being a random polynomials in $\bbS$ and $h_l \sim \ccalN(0,1)$; and 2) $\bbC_{MRF}=(\sigma\bbI+\delta\bbS)^{-1}$ with $\sigma$, $\delta >0$ selected to guarantee that $\bbC_{MRF}^{-1}$ is positive definite.
Finally, three different algorithms are compared: (i) ``GL'', which assumes that the signals are Gaussian in the graph; (ii) ``GSR'', which assumes that the signals are stationary in the graph; and (iii) ``GGSR'', which assumes both stationarity and Gaussianity for the graph signals.

\vspace{0.05cm}
\noindent \textbf{Performance for limited number of samples.}
In this experiment, we quantify the recovery performance of the different algorithms as the number of signals $R$ increases (left panel of Fig. \ref{F:num_exp}). 
Moreover, we consider the two covariance models mentioned above: $\bbC_{MRF}$ and $\bbC_{poly}$. 
Looking at the results obtained when using the $\bbC_{MRF}$ covariance model, we observe that GGSR provides a ratio of recovered graphs very similar to GL (which is the maximum-likelihood estimator under $\bbC_{MRF}$) and clearly outperforms GSR. 
Regarding the results when using the covariance model $\bbC_{poly}$ it can be seen that GGSR outperforms both GSR and GL, the latter being unable to estimate the topology of any graph.
In summary, when considering the $\bbC_{MRF}$ model, GL offers the best estimation because its assumption fits the $\bbC_{MRF}$ covariance model perfectly and GGSR, which incurs higher computational complexity, obtains very similar results. Differently, a big performance gap is observed under the $\bbC_{poly}$ data generation model, where GGSR achieves a recovery performance similar to that of GSR while requiring 10 times less samples. 
This experiment validates our claims (laid out in Sec. \ref{S:netw_reconstruction}) about the ability of GGSR to find a ``sweet spot'' between GL and GSR. 

\vspace{0.05cm}
\noindent \textbf{Performance for noisy samples.}
In most setups, the signals at hand may contain a certain amount of noise. 
In this test case, we evaluate the robustness of the different algorithms for learning the graph topology from a set of noisy samples, keeping $R$ constant and increasing the noise present in $\bbX$ (center panel of Fig. \ref{F:num_exp}). 
For the $\bbC_{MRF}$ model, we observe that the performance is fairly robust for all algorithms, showing only a slight deterioration as the noise increases. The two main findings from the curves associated with the $\bbC_{poly}$ model are: i) the impact of the noise is much more noticeable than for $\bbC_{MRF}$ and ii) GGSR is less sensitive to noise than GSR, with GGSR being able to correctly recover approximately twice as many graphs as GSR.

\vspace{0.05cm}
\noindent \textbf{Learning clustered graph from financial stocks.}
Finally, we test our algorithm on a dataset with real data. We selected 40 companies from 4 different sectors of the S\&P 500 index (10 companies from each sector). 
We collected the daily log returns of each company in the period Jan. 1st, 2010 to Jan. 1st, 2016  totaling 1510 days worth of  data. 
Our goal for this experiment is to estimate the graph structure from $\bbX \in \reals^{40 \times 1510}$ using GL, GSR and GGSR. 
The first column of the right panel in Fig. \ref{F:num_exp} represents the graph estimation as the thresholded version of the correlation matrix obtained from the data. 
The remaining 3 columns correspond to the estimation performed by GSR, GL and GGSR, respectively. 
The first row of the right panel in Fig. \ref{F:num_exp} represents the graph estimated by each of the 4 algorithms, with the different colors representing the sector to which each company belongs. Since we do not have a ground truth graph, to asses the quality of the learned graphs, we leverage the fact that the companies belonged to 4 financial sectors and run a clustering algorithm using as input each of the learned graphs. 
The second row of the right panel in Fig. \ref{F:num_exp} represents the output of a spectral clustering algorithm carried out over each of the graphs in the first row and setting the number of clusters to 4. 
From the results, we observe that the clustering based on the graph estimated by GGSR is the one leading to less clustering errors.
In addition, note that GGSR shows higher intra-cluster connections and lower inter-cluster connections compared to the other algorithms. 
\vspace{-0.2cm}
%%%%%%%%%%%%%%%%%%%%%%%%%%%%%%%%%%%%%%%%%%%%%%%%%%%%%%%%%%%%%%%%%%%%%%%%%%
\vspace{-0.2cm}
\section{Conclusions}\label{S:conclusions}
\vspace{-0.2cm}
This paper addressed the problem of estimating a graph from nodal observations under the assumption that the observed signals were Gaussian and graph stationary. Relative to classical Gaussian Markovian graphical models, the stationary assumption covers a wider range of scenarios where GL fails to obtain the true graph. 
%Relative to classical graph-stationary formulations, our algorithm leverages additional structure, being able to estimate correctly the graph structure with much fewer samples. 
The estimation problem was formulated as a constraint optimization where both the graph and the precision matrix of the Gaussian distribution were optimization variables, resulting in a non-convex formulation. An alternating method with convergence guarantees was designed and the recovery performance was assessed using synthetic and real data.
%%%%%%%%%%%%%%%%%%%%%%%%%%%%%%%%%%%%%%%%%%%%%%%%%%%%%%%%%%%%%%%%%%%%%%%%%%%%%%%%%%%%%%%%%%%%%%%%%%%%%%%%%%%%%%%%%%%%%%%%%%%

\vfill\pagebreak
\bibliographystyle{IEEE}
\bibliography{citations}

\end{document}